\documentclass{iau} 
\usepackage{graphicx}
\usepackage{caption}
\usepackage{subcaption}

\title[Spotless active regions]{The behavior of the spotless active regions during the solar minimum 23-24}

\author[A.~J.~Oliveira~e~Silva \& C.~L.~Selhorst]{Alexandre~Jos\'e~de~Oliveira~e~Silva$^1$ \and Caius~Lucius~Selhorst$^{1,2}$}

\affiliation{$^1$IP\&D - Universidade do Vale do Para\'iba (UNIVAP) - S\~ao Jos\'e dos Campos, SP, Brazil \\ email: {ajoliveiraesilva@gmail.com} \\[\affilskip]
$^2$NAT -€" Universidade Cruzeiro do Sul - S\~ao Paulo, SP, Brazil \\ email: {caiuslucius@gmail.com}}

\pubyear{2016}
\volume{328}  %% insert here IAU Symposium No.
\jname{Living Around Active Stars}
\editors{D. Nandi, A. Valio \& P. Petit, eds.}
\begin{document}

\maketitle

\begin{abstract}
In this work, we analysed the physical parameters of the spotless actives regions observed during solar minimum 23 -- 24 (2007 -- 2010).
The study was based on radio maps at 17~GHz obtained by the Nobeyama Radioheliograph (NoRH) and magnetograms provided by the Michelson Doppler Imager (MDI) on board the Solar and Heliospheric Observatory (SOHO). The results shows that the spotless active regions presents the same radio characteristics of a ordinary one, they can live in the solar surface for long periods ($>10$~days), and also can present small flares.

\end{abstract}

\firstsection 

\section{Introduction}

The study of the solar magnetic field dynamics  is very importante to understand the phenomena which may affect the Earth environment. These dynamics have been daily monitored since the years $\sim1600$ by the number of  sunspot (SSN). Another classical index used to measure the solar activity is the radio flux at 10.7~cm ($F10.7$), which is generated at coronal heights and related to the presence of active regions and the occurrence of flares. The relationship between these indexes, that used to be linear, was destroyed in the last two cycles, which presented a SSN smaller than the expected by the measured $F10.7$ flux (\cite[Livingston {et~al.} 2012]{Livingston2012}).

\cite{Selhorst2014} studied the number of active regions  observed by the NoRH (Nobeyama Radioheliograph) at 17~GHz between the years 1992 and 2013. During the quiet solar period between 2008 and 2009, they reported the presence of active regions during days without sunspots (about 33\% of active days in the period). This fact was addressed to regions with magnetic field intensity less than 1500~G (\cite{Livingston2012}). In this work, we investigated the physical parameters of the spotless actives regions observed during solar minimum period between 2007 and 2010.

\section{Data Analyses and Results}

In this study, the active regions at 17~GHz were identified, just as \cite{Selhorst2014}, that is: 
%\hspace*{1 cm}~~i) size greater than $150~pixels^2$ ($\sim300MSS$ (millionth solar surface));\\
i) size greater than $150~pixels^2$ ($\sim300MSS$ (millionth solar surface)); ii) latitudes between $\pm45^\circ$; iii) maximum brightness temperature ($T_{b_{max}}$) at least $40\%$ greater than the quiet Sun value and iv) longitudes $< 70^\circ$. The active region size was calculated considering the pixels with $T_b$ equal to 12000~K or more. 

During the last minimum (2007 -- 2010), our analyses find 92 days without sunspots, that presented active regions at 17~GHz. Since, an active region can live for long periods, those spotless active regions were identified as 48 distinct ones. During their lives, some of these active regions presented days with sunspot associated, these days were also analysed (128 days). These weak active regions may also present flares, and a small one (class C1.1) was identified.

\begin{figure}[!h]
   \centering
   \includegraphics[width=7.5cm]{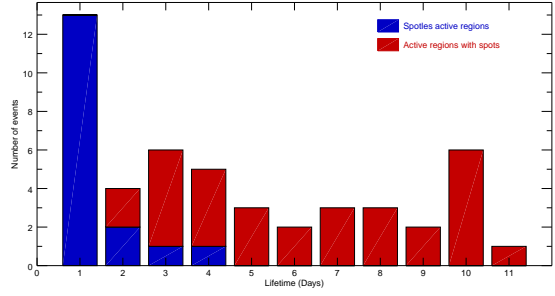} 
   \caption{The lifetime of spotless active regions (in days).  In red were represented the active regions that presented at least one day with sunspot associated.}
   \label{Fig01}
\end{figure}

The lifetimes of the spotless active regions analysed are shown in Figure \ref{Fig01}. Almost 50\% of them (23) were ephemeral ones and still visible for a maximum of three days. The results show that the spotless active regions present a short lifetime but can also live on the solar surface for long periods ($ >~10$~days). All of these active regions living 5 days or more  presented at least one day with a sunspot associated to them.

\begin{figure}[!h]
   \centering
   \includegraphics[width=7.5cm]{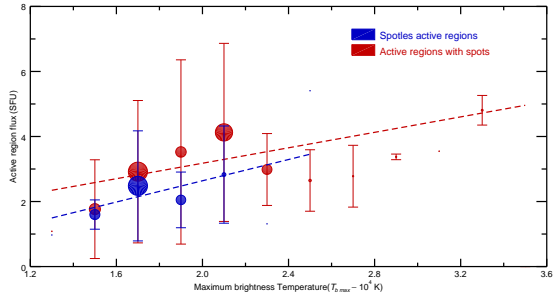} 
   \caption{The relation between the active region flux and their maximum brightness temperature ($T_{b_{max}}$). The size of circles are proportional to the percentage of active regions in each group. Moreover, the active regions were separated in with or without spot, respectively, red and blue. The dashed lines are linear adjusts.}
   \label{Fig02}
\end{figure}

The active regions presented a minimum brightness temperature of $\sim13700$~K and the maximum brightness temperature of $\sim24500$~K for the spotless days and $52600~K$  for thats ones sunspots. Moreover, the the spotless ones showed an average area of $325~pixels^2$, whereas, those with sunspot were 35\% greater ($440~pixels^2$).

In Figure~\ref{Fig02}, we compare active region flux (in SFU) with their maximum brightness temperature. The active regions were separated in groups by their $T_{b_{max}}$, every  2000~K, and the result shows the increase of the flux with $T_{b_{max}}$ for both groups the spotless regions (blue circles) and for those with associated sunspot (blue circles). The mean flux difference is 0.36 SFU, however, this difference reaches 0.85 SFU when the $T_{b_{max}}$ is lower, where the largest number of spotless regions are concentrated, and decreases when the spotless active regions reached their maximum values ($\sim 25000$~K). 

\begin{figure}[!h] 
   \centering
   \includegraphics[width=13.5cm]{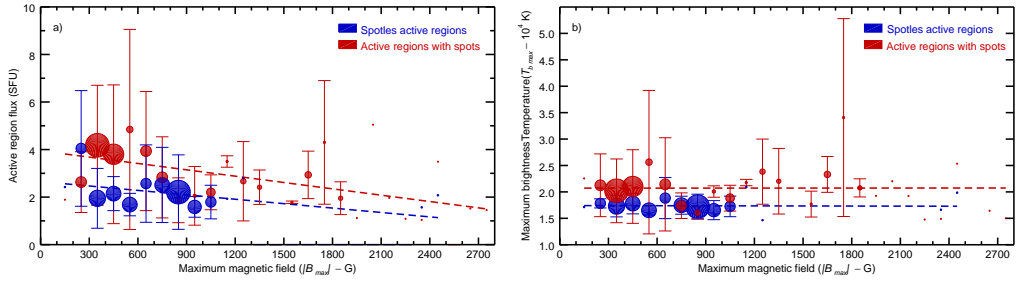} 
   \caption{Comparison of the (a) active region flux (SFU) and (b) mean brightness temperature in relation to $|B|_{max}$.}
   \label{figlinear}
\end{figure}

To analise the active regions magnetic fields intensities, the magnetograms obtained by the MDI (Michelson Doppler Imager)  were analysed. Due to sight line, the values of the maximum intensities of magnetic fields ($|B|_{max}$) were corrected by dividing the value obtained in the magnetogram by $(\cos(Lat.)\times\cos(Long.))$ (\cite{Schad_Penn2010}). Here, we used the absolute maximum magnetic field intensity ($|B|_{max}$) to characterise the active regions. 

In the Figure~\ref{figlinear}, the active regions were grouped by their $|B|_{max}$ each 100~G. The spotless regions are plotted in blue and the active regions with spots in red. For each group, the maximum brightness temperature ($T_{b_{max}}$) and the active region flux were averaged for each group. 
The panel \ref{figlinear}(a) shows a negative trend in the linear adjust for both, active regions with spots and without them, that is, as magnetic fields increase the flux tends to decrease.  The averaged flux of the active regions with spots is 0.81 SFU greater than the spotless ones. In panel \ref{figlinear}(b), $T_{b_{max}}$ still constant with the increase of the magnetic field. Moreover, the active regions with spots are $3400$~K hotter than the spotless ones.

\section{Final Remarks}
A total of 48 distinct active regions were analysed in the period 2007--2010. About 50\% of them were ephemeral living a maximum of three days. On the other hand, those ones living 5 days more presented at least one day with a sunspot. The active regions with sunspots are hotter and presented more flux than the spotless ones. However, the values were significantly smaller than the proposed by Livingston et~al. (2012) for minimum necessary for the spot formation (1500~G), that could be due to instrumental differences.

\begin{acknowledgements}
 We would like to thank the Nobeyama Radioheliograph, which is
operated by the NAOJ/Nobeyama Solar Radio Observatory. A.J.O.S. acknowledge the scholarship form CAPES. C.L.S. acknowledge financial support from the S\~ao Paulo Research Foundation (FAPESP), grant  \#2014/10489-0. 
\end{acknowledgements}

\end{document}